\documentclass[preprint2]{aastex6}
\usepackage{comment}

\shorttitle{A \tess Earth-sized Planet and a Sub-Neptune}
\shortauthors{Dragomir et al.}

\newcommand{\exofast}{{\it EXOFASTv2 }} 
\newcommand{\tess}{{\it TESS }}

\providecommand{\bjdtdb}{\ensuremath{\rm {BJD_{TDB}}}}

\providecommand{\msun}{\ensuremath{\,M_\Sun}}
\providecommand{\rsun}{\ensuremath{\,R_\Sun}}
\providecommand{\lsun}{\ensuremath{\,L_\Sun}}

\def\smallskip{\vspace\smallskipamount}
\newskip\smallskipamount \smallskipamount=0.5pt plus 0pt minus 1.5pt

\begin{document}





\bgroup
\renewcommand*{\thefootnote}{\fnsymbol{footnote}}
\footnote{This paper includes data gathered with the 6.5 meter Magellan Telescopes located at Las Campanas Observatory, Chile.}
\egroup
\title{TESS delivers its first Earth-sized planet and a warm sub-Neptune$^*$}

\email{dragomir@space.mit.edu}

\author{Diana Dragomir\altaffilmark{1,2}, 
Johanna Teske\altaffilmark{2,3,4},
Maximilian N.\ G{\"u}nther\altaffilmark{1,5},
Damien S\'egransan\altaffilmark{6}, 
Jennifer A. Burt\altaffilmark{1,5},
Chelsea X. Huang\altaffilmark{1,5},
Andrew Vanderburg\altaffilmark{7,8},
Elisabeth Matthews\altaffilmark{1},
Xavier Dumusque\altaffilmark{6},
Keivan G. Stassun\altaffilmark{9},
Joshua Pepper\altaffilmark{10},
George R. Ricker\altaffilmark{1},
Roland Vanderspek\altaffilmark{1},
David W. Latham\altaffilmark{11}, 
Sara Seager\altaffilmark{1,12,13}, 
Joshua N. Winn\altaffilmark{14}, 
Jon M. Jenkins\altaffilmark{15},
Thomas Beatty\altaffilmark{16,17}, 
Fran\c cois Bouchy\altaffilmark{6}, 
Timothy M. Brown\altaffilmark{18,19},
R. Paul Butler\altaffilmark{3}, 
David R. Ciardi\altaffilmark{20}, 
Jeffrey D. Crane\altaffilmark{4}, 
Jason D. Eastman\altaffilmark{11}, 
Luca Fossati\altaffilmark{21}, 
Jim Francis\altaffilmark{1}, 
Benjamin J. Fulton\altaffilmark{20}, 
B. Scott Gaudi\altaffilmark{22}, 
Robert F. Goeke\altaffilmark{1}, 
David James\altaffilmark{11}, 
Todd C. Klaus\altaffilmark{23}, 
Rudolf B.\ Kuhn\altaffilmark{24,25}, 
Christophe Lovis\altaffilmark{6}, 
Michael B. Lund\altaffilmark{9}, 
Scott McDermott\altaffilmark{26}, 
Martin Paegert \altaffilmark{11}, 
Francesco Pepe \altaffilmark{6}, 
Joseph E. Rodriguez\altaffilmark{11}, 
Lizhou Sha\altaffilmark{1},
Stephen A. Shectman\altaffilmark{4}, 
Avi Shporer\altaffilmark{1},
Robert J. Siverd\altaffilmark{9}, 
Aylin Garcia Soto \altaffilmark{12}, 
Daniel J. Stevens\altaffilmark{16,17}, 
Joseph D. Twicken\altaffilmark{15,27}, 
St{\'e}phane Udry\altaffilmark{6}, 
Steven Villanueva Jr.\altaffilmark{1}, 
Sharon X. Wang\altaffilmark{3}, 
Bill Wohler\altaffilmark{15,27}, 
Xinyu Yao\altaffilmark{10}, 
Zhuchang Zhan\altaffilmark{1,12}}

\altaffiltext{1}{Department of Physics and Kavli Institute for Astrophysics and Space Research, Massachusetts Institute of Technology, Cambridge, MA 02139, USA}
\altaffiltext{2}{NASA Hubble Fellow}
\altaffiltext{3}{Department of Terrestrial Magnetism, Carnegie Institution for Science, 5241 Broad Branch Road, NW, Washington, DC 20015, USA}
\altaffiltext{4}{Observatories of the Carnegie Institution for Science, 813 Santa Barbara Street, Pasadena, CA 91101, USA}
\altaffiltext{5}{Juan Carlos Torres Fellow}
\altaffiltext{6}{Astronomy Department of the University of Geneva, 51 chemin des Maillettes, 1290 Versoix, Switzerland}
\altaffiltext{7}{Department of Astronomy, The University of Texas at Austin, Austin, TX 78712, USA}
\altaffiltext{8}{NASA Sagan Fellow}
\altaffiltext{9}{Department of Physics \& Astronomy, Vanderbilt University, 6301 Stevenson Center Ln., Nashville, TN 37235, USA}
\altaffiltext{10}{Department of Physics, Lehigh University, 16 Memorial Drive East, Bethlehem, PA 18015, USA}
\altaffiltext{11}{Center for Astrophysics $|$ Harvard \& Smithsonian, 60 Garden Street, Cambridge, MA 01238, USA}
\altaffiltext{12}{Department of Earth, Atmospheric and Planetary Sciences, MIT, 77 Massachusetts Avenue, Cambridge, MA 02139, USA}
\altaffiltext{13}{Department of Aeronautics and Astronautics, MIT, 77 Massachusetts Avenue, Cambridge, MA 02139, USA}
\altaffiltext{14}{Department of Astrophysical Sciences, Princeton University, 4 Ivy Lane, Princeton, NJ 08544, USA}
\altaffiltext{15}{NASA Ames Research Center, Moffett Field, CA, 94035, USA} 
\altaffiltext{16}{Department of Astronomy \& Astrophysics, The Pennsylvania State University, 525 Davey Lab, University Park, PA 16802, USA}
\altaffiltext{17}{Center for Exoplanets and Habitable Worlds, The Pennsylvania State University, 525 Davey Lab, University Park, PA 16802, USA}
\altaffiltext{18}{Las Cumbres Observatory, 6740 Cortona Dr., Suite 102, Goleta, CA 93117, USA}
\altaffiltext{19}{University of Colorado/CASA, Boulder, CO 80309, USA}
\altaffiltext{20}{Caltech/IPAC-NExScI, 1200 East California Boulevard, Pasadena, CA 91125, USA}
\altaffiltext{21}{Space Research Institute, Austrian Academy of Sciences, Schmiedlstrasse 6, A-8042 Graz, Austria}
\altaffiltext{22}{The Ohio State University, Department of Astronomy, Columbus, OH 43210, USA}
\altaffiltext{23}{Stinger Ghaffarian Technologies, Moffett Field, CA, 94035, USA}
\altaffiltext{24}{South African Astronomical Observatory, PO Box 9, Observatory, 7935, Cape Town, South Africa}
\altaffiltext{25}{Southern African Large Telescope, PO Box 9, Observatory, 7935, Cape Town, South Africa}
\altaffiltext{26}{Proto-Logic LLC, 1718 Euclid Street NW, Washington, DC 20009, USA}
\altaffiltext{27}{SETI Institute, Mountain View, CA 94043, USA}

\begin{abstract}

The future of exoplanet science is bright, as \tess once again demonstrates with the discovery of its longest-period confirmed planet to date. We hereby present HD 21749b (TOI 186.01), a sub-Neptune in a 36-day orbit around a bright (V = 8.1) nearby (16 pc) K4.5 dwarf. \tess measures HD21749b to be 2.61$^{+0.17}_{-0.16}$ $R_{\oplus}$, and combined archival and follow-up precision radial velocity data put the mass of the planet at $22.7^{+2.2}_{-1.9}$ $M_{\oplus}$. HD 21749b contributes to the \tess Level 1 Science Requirement of providing 50 transiting planets smaller than 4 $R_{\oplus}$ with measured masses. Furthermore, we report the discovery of HD 21749c (TOI 186.02), the first Earth-sized ($R_p = 0.892^{+0.064}_{-0.058} R_{\oplus}$) planet from \tess. The HD21749 system is a prime target for comparative studies of planetary composition and architecture in multi-planet systems.

\end{abstract}


\section{Introduction}

Small exoplanets are common in the Milky Way \citep{Howard2012, Fressin13, Ful17}, but for a long time astronomers have had an incomplete picture of their properties. The recently-launched Transiting Exoplanet Survey Satellite ({\it TESS}) is revolutionizing the field of exoplanet science by discovering planets of all sizes around the nearest stars. The mass, atmospheric composition and other previously mostly inaccessible properties of small exoplanets will be measurable for many \tess systems. Four \tess-discovered planets smaller than Neptune have already been announced. $\pi$ Men c is a 2 $R_{\oplus}$ super-Earth transiting its naked-eye G0V star every 6.3 days \citep{Hua18b}. Since its mass is measured, $\pi$ Men c contributes toward the \tess Level 1 Science Requirement of providing 50 transiting planets smaller than 4 $R_{\oplus}$ with measured masses \citep{Ric15}. LHS 3844b is a 1.3 $R_{\oplus}$ hot terrestrial planet, orbiting its M4 dwarf star every 11 hours \citep{Van19}. The last two (TOI 125b and c; \citealt{Qui19}) are statistically validated sub-Neptunes orbiting a K0 dwarf with periods of 4.65 and 9.15 days. All of these discoveries are based on only the first two sectors of \textit{TESS} data, suggesting many more are to be found.

Longer-period transiting planets are notoriously difficult to find because their transit probability is lower and even if they do transit, they do so less frequently. We know of very few around nearby stars. Only three small ($R_p < 4~R_{\oplus}$) planets with measured masses are known to have orbital periods greater than 5 days and transit stars with V-band magnitudes brighter than 10: $\pi$ Men c \citep{Hua18b}, HD 219134c \citep{Gil17} and HD 97658b \citep{Dra13}, the latter having the longest period of the three, at 9.5 days. Most of the exoplanets that \tess will reveal will have orbital periods shorter than 10 days \citep{Sul15, Bar18, Hua18}. In most parts of the sky \tess may only observe one or two transits for longer-period planets \citep{Vil19}, making them more challenging to detect and confirm, particularly for small planets.

In this Letter we present HD 21749b (\tess Object of Interest 186.01), a sub-Neptune with a period of 35.61 days that initially appeared as a single-transit planet candidate. We also introduce a second planet in the same system (HD 21749c), with a period of 7.8 days. The host star is a bright (V = 8.1) K dwarf, located only 16 pc away, making this system likely to satisfy the follow-up interests of many exoplanet astronomers. In Section \ref{sec:obs} we describe the \tess photometry and the other observations used to confirm HD 21749b and validate HD 21749c. In Section \ref{sec:analysis} we describe our analysis and results. We discuss the implications of our findings and conclude in Section \ref{sec:conc} \footnote{An independent analysis of archival HARPS RVs of HD 21749 has been reported by \citet{Tri19}. Our paper differs in scope in that we confirm HD 21749b using \tess photometry, HARPS and PFS RVs, and we report an additional planet in the system.}

\section{Observations}
\label{sec:obs}

\subsection{\tess Photometry}
\label{sec:tessphot}

\tess will survey nearly the entire sky over two years by monitoring contiguous overlapping 90 $\times$ 24$^{\circ}$ sectors for 27 days at a time \citep{Ric15}. The primary mission will complete the southern ecliptic hemisphere in its first year, and the northern hemisphere in its second. Toward the ecliptic poles (i.e. higher ecliptic latitudes), there is overlap between sectors and targets can be observed for more than 27 days. This is the case for HD 21749, which was observed in four \tess sectors. We used the publicly available four-sector \tess data in our analysis.

\begin{figure*}[!ht]
    \centering
    \includegraphics[width=2\columnwidth]{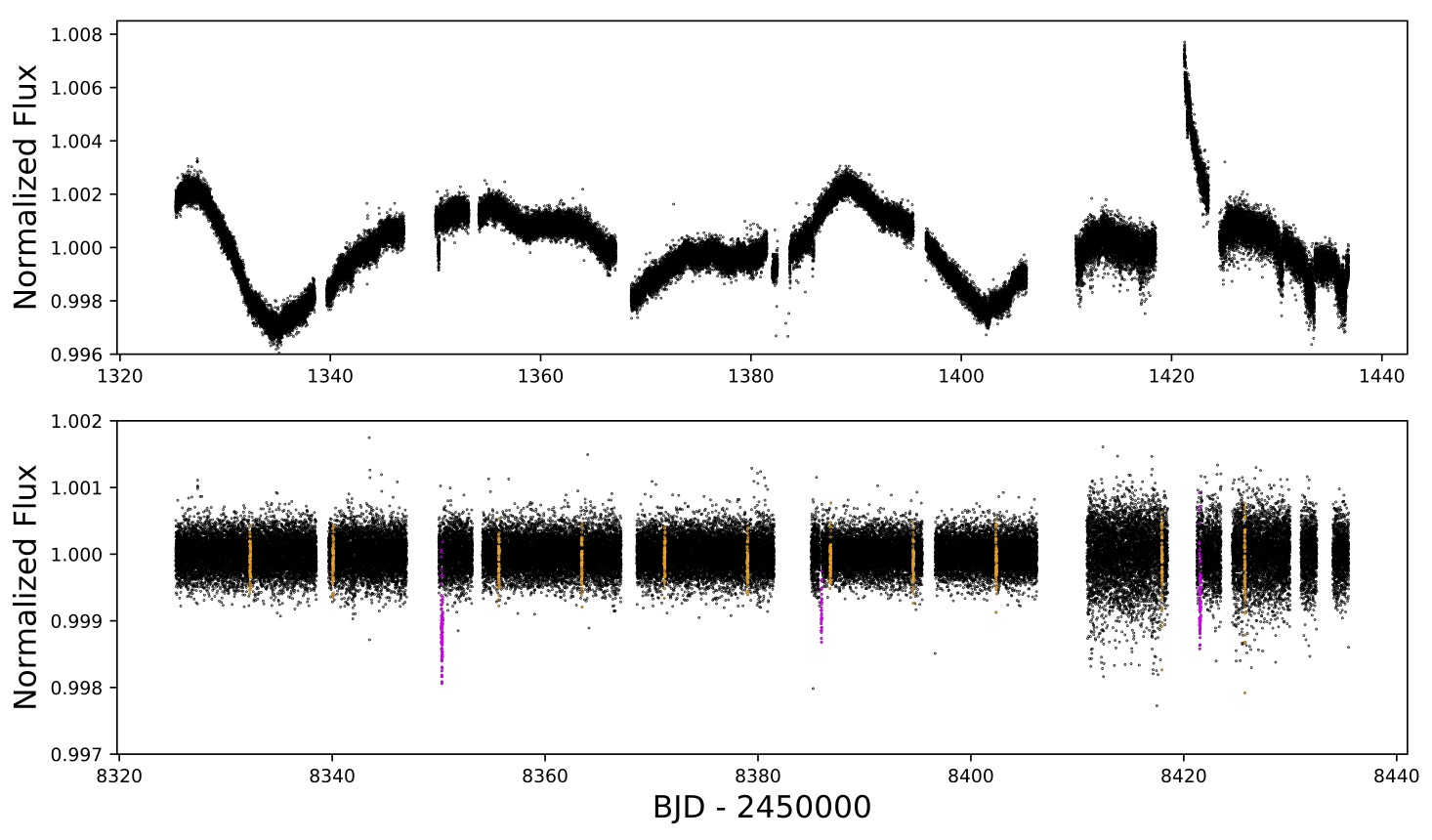}
    \includegraphics[width=2\columnwidth]{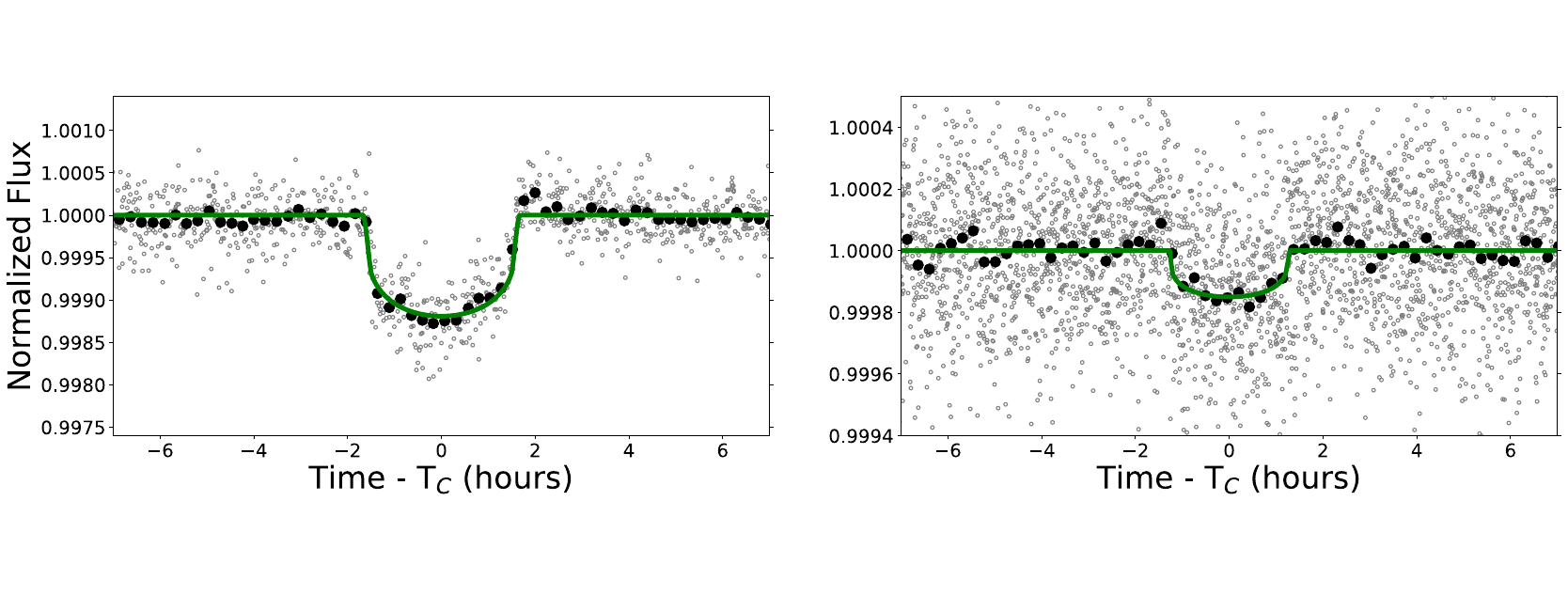}
    \caption{\tess photometry of HD 21749. {\it Top:} Raw light curve spanning four sectors. An interruption in communications occurred between the instrument and the spacecraft between 1418.54 and 1421.21. It was followed by a temperature increase due to heaters being turned on. Once data collection resumed, the heaters were turned off, but the camera temperature took about three days to return to nominal, causing the sharp feature in flux seen between 1421 and 1424. This effect was easily removed by our detrending procedure (see Sections \ref{sec:tessphot} and \ref{sec:fit} for details. {\it Middle:} Detrended light curve. Transits of HD 21749b and HD 21749c are shown in magenta and orange, respectively. {\it Bottom:} Phase-folded transits of HD 21749b (left) and HD 21749c (right).  In both plots, the binned light curves and best-fit transit models are plotted with black points and green lines, respectively.}
    \label{fig:tess}
\end{figure*}

The first transit of HD 21749b was identified by both the MIT Quick Look Pipeline (which searches for planet candidates in the 30-min Full Frame Images) and the Science Processing Operations Center (SPOC) pipeline based at the NASA Ames Research Center \citep{Jen2016}. No other matching transits were found in the publicly-released data from sectors 1 and 2. After TOI 186.01 was alerted, we searched for archival spectroscopy of this very bright star and found 59 HARPS radial velocities (RVs) in the ESO archive (see \ref{sec:harps}). A periodogram of these RVs showed a clear signal at 35.57 days, but the \tess photometry and the R$_{HK}$ index \citep{Bor18} indicate a stellar rotation period of around 35 days, calling for caution. If the strongest period in the RVs did correspond to the planet, then we expected to see additional transits in sectors 3 and 4. Once the sector 3 data were released, we discovered that a momentum dump \footnote{``Momentum dumps" consist of resetting the momentum wheel speed every 2.5 - 3 days and are used to mitigate the noisier-than-expected measurements of the spacecraft momentum wheel speeds at higher wheel speeds (see \url{https://archive.stsci.edu/files/live/sites/mast/files/home/missions-and-data/active-missions/tess/_documents/TESS_Instrument_Handbook_v0.1.pdf} for details). Momentum dumps require brief interruptions to Fine Pointing mode, during which an increase in the flux dispersion is noticeable in the science data, so data acquired during these intervals are excluded from our analysis.} occurred approximately 35.6 days after the sector 1 transit (see Figure \ref{fig:tess}). We did not let this unexpected turn of events foil our search efforts, and upon close inspection of the light curve we succeeded in recovering a partial transit (including egress) immediately following the momentum dump. Finally, we observed a third transit in sector 4, thus allowing for a robust ephemeris determination (see Section \ref{sec:fit}). Serendipitously, when applied to the first three sectors of the HD 21749 light curve, the SPOC Pipeline yielded an additional planet candidate with a period of 7.9 days (TOI 186.02).

We used the 2-minute target pixel data for our analysis. The target was on the edge of the camera, where the point spread function is triangular in shape. We tried improving the light curve precision by extracting light curves from the publicly available target pixel stamps using different photometric apertures (circles as well as irregular pixel boundaries). We then detrended the raw light curve by fitting a basis spline with knots spaced by 0.3 days, after excluding both 3$\sigma$ outliers and data obtained during and immediately surrounding transits \citep{Van14}. The final 2-minute cadence light curve has an RMS of 240 ppm. 

Care must be taken to rule out false positives that could masquerade as planet candidates. Both planet candidates in the HD 21749 system pass the false positive tests performed on the \tess photometry: there is no evidence of secondary eclipses, nor any detectable motion of the centroid of the star on the detector during the transit events. While a giant star (HIP 16068/TIC 279741377; $R_*$ = 3.15$^{+0.12}_{-0.09}$ $R_{\odot}$) is present 22'' from HD 21749, the excellent period match between the transits and the HD 21749 RVs rule out this neighboring star as the source of the TOI 186.01 transit signals. If TOI 186.02 transits the giant star, it would have to be a Jupiter-sized object since the transit events have a low impact parameter (and are thus unlikely to be due to a grazing eclipsing binary). The probability that the giant star blended with HD 21749 would coincidently {\it also} host a transiting planet is very low, but we nevertheless explored this scenario in Section \ref{sec:HD21749c}. The presence of the neighboring star also warrants correcting the transit depths for dilution (see Section \ref{sec:fit}). 

\subsection{VLT NaCo Imaging}

To test for nearby stellar companions which could dilute the light curve and bias the measured planetary radius, we used archival data from the VLT/NaCo instrument \citep{Len03, Rou03} collected in 2005. These data were initially collected to search for planets and low mass stellar companions, and as such used common direct imaging techniques to maximise the achievable contrast. Images were collected with a 6s exposure time, causing the central pixels of the star to saturate. Beyond ~50mas, where the images are not saturated, the longer integration time increases the observing efficiency and reduces read noise. To complement these saturated images, a small number of images with a shorter exposure (1.9s) were collected in which the target star is unsaturated and its flux can thus be calibrated. In addition, images were collected simultaneously at multiple wavelengths, and sequentially at multiple telescope rotations, such that variations in the PSF could be used to separate stellar speckle noise and companion flux \citep{Mar06}. For simplicity, in this work we used only the 1.6$\mu$m data in the upper left quadrant of the chip, and simply co-added the aligned data. Although this reduces the absolute contrast reached, it is sufficient for the purposes of this analysis.

\begin{figure}[!ht]
    \centering
    \includegraphics[width=\linewidth,trim=0 0 0 0,clip]{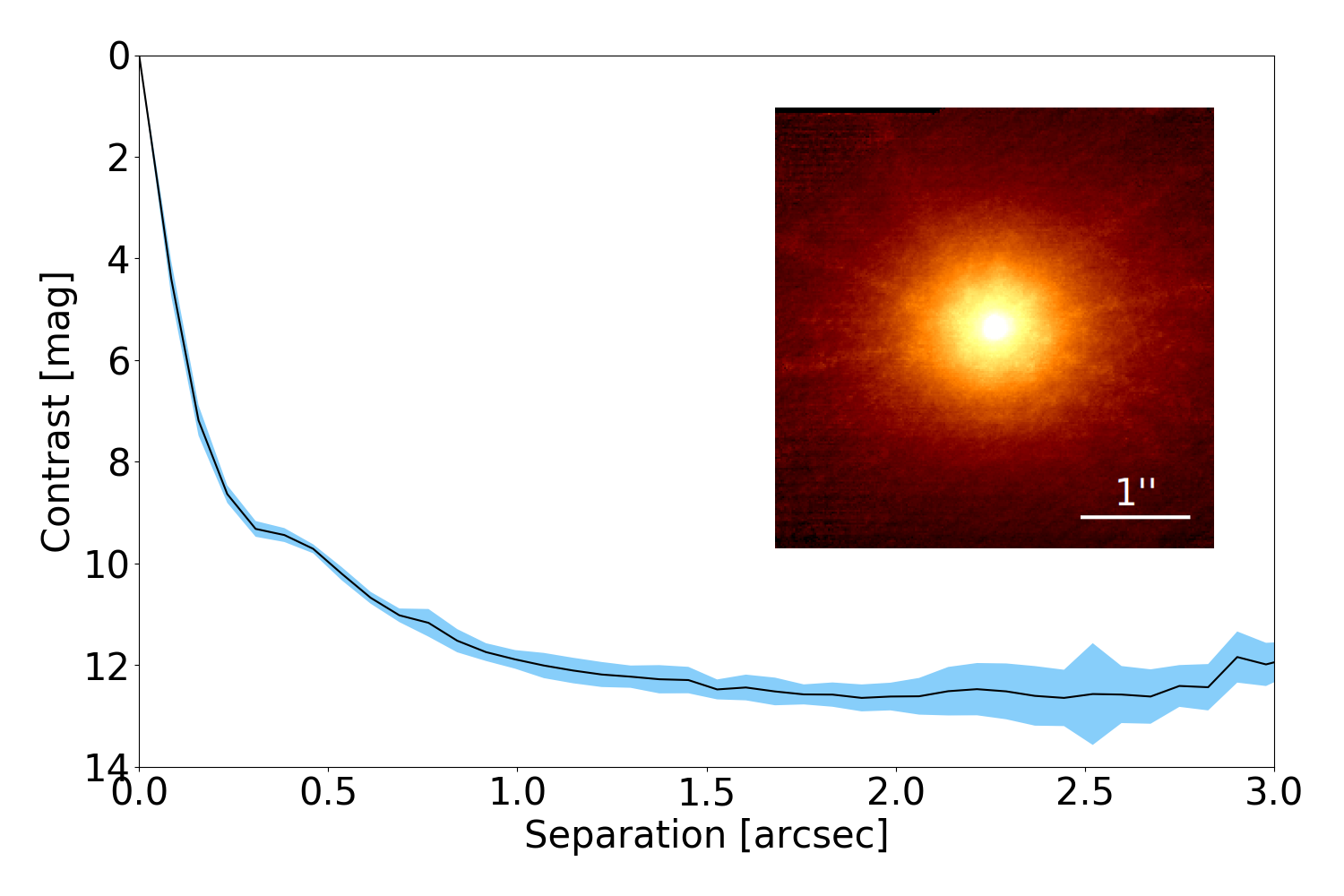}
    \caption{Sensitivity curve for the high-resolution AO imaging of the target with VLT/NaCo. The inset image is 4''$\times$4''. No companions are detected within the field of view, and the target appears single to the limit of the imaging resolution.}
    \label{fig:vltnaco}
\end{figure}

We used the first set of ten saturated frames collected on 2005-11-29 (each 28$\times$6s) -- since the conditions degrade in the second set of ten frames -- and two unsaturated frames (16$\times$1.8s) taken immediately before and after this. We followed a standard data reduction process: we subtracted a dark/sky background, cropped the data to the relevant part of the chip, flat-fielded, aligned images based on the stellar position, and derotated frames to align the north image before combining the frames into a single reduced image. To determine the sensitivity of this image, we inserted scaled copies of the unsaturated PSF, and recovered these with standard aperture photometry. A sensitivity curve and an image of the target are shown in Figure \ref{fig:vltnaco}. The target appears single to the resolution of these images, with no companions detected within the field of view, which extends to at least 2'' from the target in every direction. Beyond this separation \textit{Gaia} is sensitive to companions $\sim$6 mag fainter than the host, and a companion of this magnitude would change the measured planet radius by 0.2\%. We conclude that the measured radii of HD 21749b and HD 21749c are unbiased by nearby stars. 

\subsection{LCO-NRES and CORALIE Spectroscopy}

We used the Las Cumbres Observatory (LCO; \citealt{Bro13}) Network of Echelle Spectrographs (NRES; \citealt{Siv16}) to obtain two spectra of the neighboring star TIC 279741377, and analyzed them with SpecMatch \citep{Pet17}. The two spectra had a S/N of 59 and 69 and show RVs consistent at the 100 m/s level, thus ruling out an eclipsing binary on TIC 279741377.

We obtained additional spectroscopy with the CORALIE spectrograph \citep{Que00} on the Swiss Euler 1.2m telescope at La Silla Observatory in Chile. The spectra were collected at phases 0.18 and 0.82, near the expected radial velocity maximum and minimum. Cross-correlations were performed with a weighted K5 binary mask from which telluric and interstellar lines were removed \citep{Pep02}.

\begin{figure}[!ht]
    \centering
    \includegraphics[width=\columnwidth]{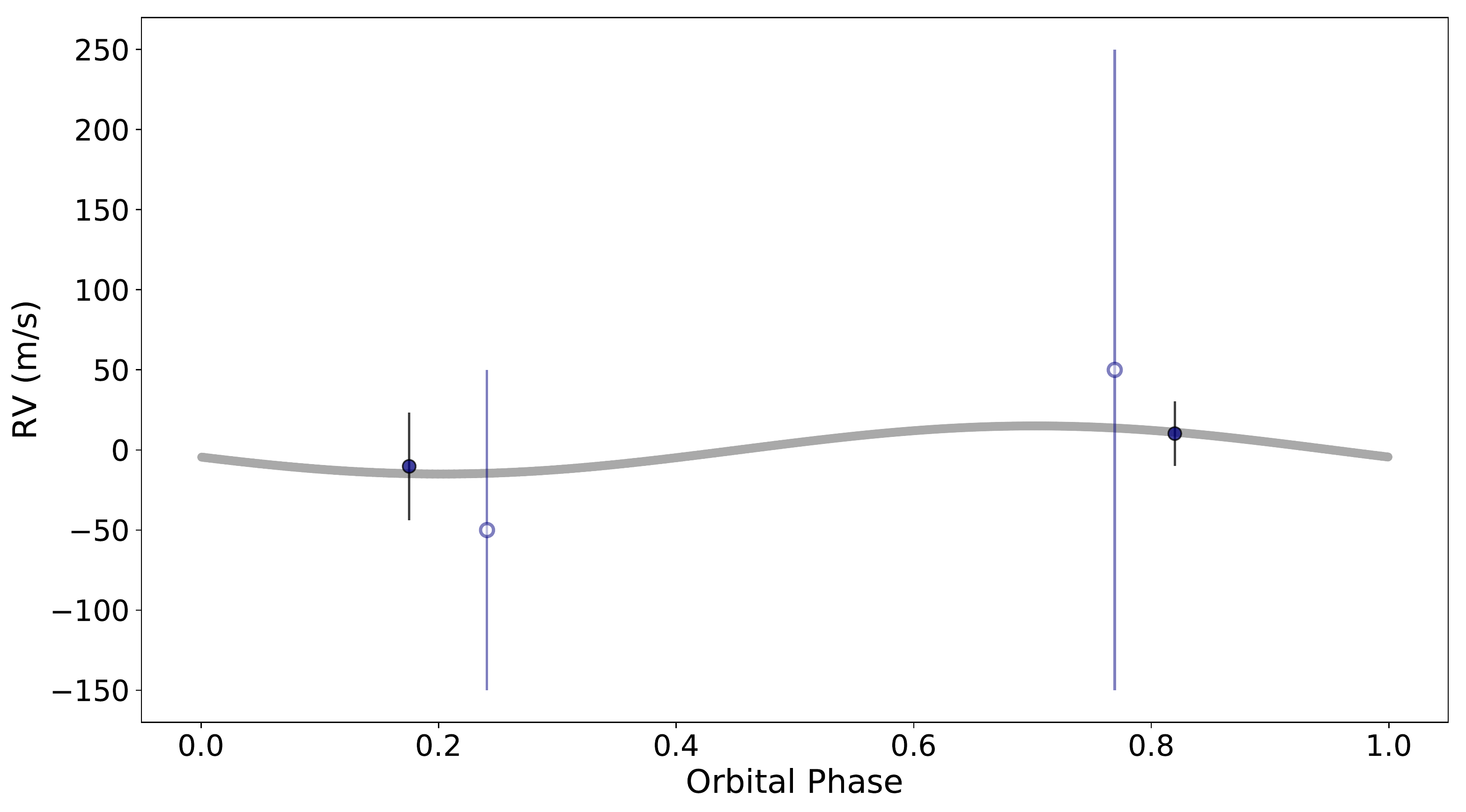}
    \caption{NRES (open circles) and CORALIE (filled circles) RVs for TIC 279741377.}
    \label{fig:nrescoralie}
\end{figure}

The NRES and CORALIE RVs are shown in Figure \ref{fig:nrescoralie}.

\subsection{HARPS Spectroscopy}
\label{sec:harps}

Prior to this work, we had collected 59 observations of HD 21749 with the High-Accuracy Radial-velocity Planet Searcher HARPS \citep{May03} on the ESO 3.6m telescope at La Silla Observatory in Chile. Of these, 55 were obtained between 2003 and 2009 by the original HARPS guaranteed time observation (GTO) programs to search for planets, and four measurements were obtained by another program in 2016. These observations are publicly available on the ESO Science Archive Facility. 

We extracted precision radial velocities from these $R\sim$115,000 spectra and utilized them to both rule out an eclipsing binary as the origin of the transit signal, and to measure the mass of HD 21749b. The data taken in 2016 have slightly different coordinates and proper motion, and the RVs were extracted using a template for a different spectral type compared to the data obtained by the HARPS GTO. We homogenized all the observations to account for these differences, and re-reduced all the data using the latest HARPS pipeline\footnote{The consistently-reduced set of HARPS RVs can be obtained through the DACE platform: dace.unige.ch.}. In addition, the 2016 observations were gathered after an upgrade to the instrument that involved a change of some optical fibres \citep{LoCurto-2015}. This modification induced a few m/s offset in RV that is dependent on the stellar spectral type. It is difficult to model this offset, and the best option so far is to fit for an offset between the HARPS data taken before and after this upgrade (see \ref{sec:fit}).

\subsection{PFS Spectroscopy}
\label{sec:pfs}

The second PRV data set presented in this work comes from the iodine-fed Planet Finder Spectrograph \citep{Cra10} on the 6.5m Magellan II telescope at Las Campanas Observatory in Chile. HD 21749 was observed with PFS, with somewhat irregular sampling, as part of the long-term Magellan Planet Search Program between January 2010 and October 2018, for a total of 48 radial velocities (45 epochs). We then began a high-cadence observing campaign purposely for \tess follow-up in December 2018, adding 34 more velocities (nine more epochs). The iodine data prior to 2018 were taken through a 0.5\arcsec slit resulting in $R\sim 80,000$, and those from 2018 were taken through a 0.3\arcsec slit, resulting in $R\sim 130,000$; the iodine-free template was taken through the 0.3\arcsec slit. The PFS detector and observing strategy changed in Feb. 2018, and the jitter in the 2018 observations decreased significantly. Exposure times ranged from $\sim$150s to $\sim$600s, resulting in S/N of $\sim$100 to 200. All PFS data are reduced with a custom IDL pipeline that flat fields, removes cosmic rays, and subtracts scattered light. Further details about the iodine-cell RV extraction method can be found in \cite{But96}, and the PFS RVs used in this paper can be found on ExoFOP-TESS\footnote{https://exofop.ipac.caltech.edu/tess/target.php?id=279741379}.

\section{Analysis and Results}
\label{sec:analysis}

\subsection{Stellar Parameters}
\label{sec:stellar}

We performed a fit to the broadband spectral energy distribution (SED), following the methodology described in \citet{Sta16}. We adopted the fluxes published in all-sky photometric catalogs: {\it Tycho-2\/} $B_T V_T$, {\it 2MASS\/} $JHK_S$, and {\it WISE}1--4. These flux measurements span the wavelength range 0.4--22~$\mu$m. We assumed solar metallicity based on values in the PASTEL catalog \citep{Sou16}, and we fit Kurucz atmosphere models \citep{Kur13} with the free parameters being the effective temperature, the overall flux normalization, and the extinction, the latter limited to the maximum line-of-sight value from the dust maps of \citet{Sch98}. The resulting SED fit has a reduced $\chi^2 = 3.4$, $T_{\rm eff} = 4640 \pm 100$~K, $A_V = 0.14^{+0.00}_{-0.04}$, $F_{\rm bol} = 2.43 \pm 0.11 \times 10^{-8}$ erg~s$^{-1}$~cm$^{-2}$. With the {\it Gaia\/} DR2 parallax \citep{Gai18} corrected for systematic offset from \citet{Sta18b}, this gives $R_\star = 0.695 \pm 0.030~R_\odot$. 
Note that we have used here the simple $1/\pi$ distance estimate with symmetric errors; in this case the parallax is so large and the relative uncertainty so small that this simple estimate is nearly identical to the Bayesian estimate with asymmetric errors \citep{Bai18}.
We can estimate the stellar mass from the empirical relations of \citet{Tor10}, which gives $M_\star = 0.73 \pm 0.07~M_\odot$. Together with the stellar radius, this provides an empirical estimate of the stellar mean density, $\rho_\star = 3.09 \pm 0.23$ g~cm$^{-3}$. 

We also derived the stellar age, employing the PARAM online Bayesian interface \citep{Das06}, which interpolates the apparent V-band magnitude, parallax, effective temperature, and metallicity onto PARSEC stellar evolutionary tracks \citep{Bre12}. We employed the initial mass function from \citet{Cha01} and a constant star formation rate, obtaining an age of 3.8 $\pm$ 3.7 Gyr.

The top section of Table \ref{tab:pars} lists the stellar parameters.

\subsection{Validation of HD 21749c}
\label{sec:HD21749c}

In this Section we rule out the remaining false positive scenario for HD 21749c. The non-grazing impact parameter of the TOI 186.02 transits and the NRES and CORALIE RVs of TIC 279741377 (Figure \ref{fig:nrescoralie}) rule out that the candidate is an eclipsing binary on this star. If TOI 186.02 were a planet transiting TIC 279741377, it would have a radius of 1.3 R$_{J}$, for which the mass-radius relations of \cite{Ning2018} suggest a mass of $\approx$ 1 M$_{J}$. We used the $T_{eff}$, $[Fe/H]$ and log$g$ values from the SpecMatch analysis of the NRES spectra together with isochrone modeling (as described in \citealt{Joh17}) to obtain a stellar mass for TIC 279741377 of 1.86 $pm$ 0.17 M$_{\odot}$. We determined that a 1 M$_J$ planet around this star would result in a RV semi-amplitude of 65 m/s, which is ruled out by the CORALIE RVs. Moreover, the equatorial transit duration in this scenario (assuming a circular orbit) would be 9.2 hours -- 3.7 times longer than the transit of TOI 186.02. To match the transit duration of TOI 186.02, the planet would have to transit TIC 279741377 with an impact parameter of 0.96, which is inconsistent with the value we find for TOI 186.02.

\subsection{Joint Photometry and Radial Velocity Fit}
\label{sec:fit}
We further detrended the spline-corrected light curves (described in Section 2.1) by training a Gaussian Process (GP; \citealt{Foreman-Mackey2017}) on the out-of-transit data, and subsequently evaluating it on the full lightcurve \citep{Gue17}. We conservatively discarded the last transit of HD 21749c because it occurred during a momentum dump. We then performed two independent analyses that jointly fitted the \tess photometry with the HARPS and PFS RVs, using a 2-planet model with a long-term trend ($\dot{\gamma}$). We used \exofast \citep{Eas17,Eas13} and {\it allesfitter} (G{\"u}enther \& Daylan, in prep.). \exofast is based on a differential evolution Markov Chain Monte Carlo algorithm that uses error scaling. {\it allesfitter} is an inference framework that unites the packages \texttt{ellc} \citep[light curve and RV models;][]{Max16}, \texttt{dynesty} (static and dynamic nested sampling; \url{https://github.com/joshspeagle/dynesty}),
\texttt{emcee} \citep[MCMC sampling;][]{DFM13} and 
\texttt{celerite} \citep[GP models;][]{Foreman-Mackey2017} to model systematic noise in the photometry. We used \texttt{allesfitter} with nested sampling instead of MCMC.

We fitted for constant offsets and jitter terms (added in quadrature to the instrumental uncertainties) for each of the four RV data sets we used: HARPS 1 (pre-upgrade), HARPS 2 (post-upgrade), PFS 1 (pre-upgrade), PFS 2 (post-upgrade). We used the stellar mass, radius, metallicity and temperature values listed in \ref{sec:stellar} as Gaussian priors for both analyses. We also fit for dilution of the transit signal due to the neighboring star, for which we used the contamination ratio from TIC V7 \citep{Sta18c} and 10\% of its value as a Gaussian prior. We did not use any other Gaussian priors. We note that \exofast determines and uses quadratic limb darkening coefficients interpolated from the Claret tables for \tess \citep{Cla17}, while {\it allesfitter} uniformly samples the quadratic limb darkening coefficients following \citet{Kipping2013}.

\begin{figure*}[!ht]
    \centering
    \includegraphics[width=2\columnwidth]{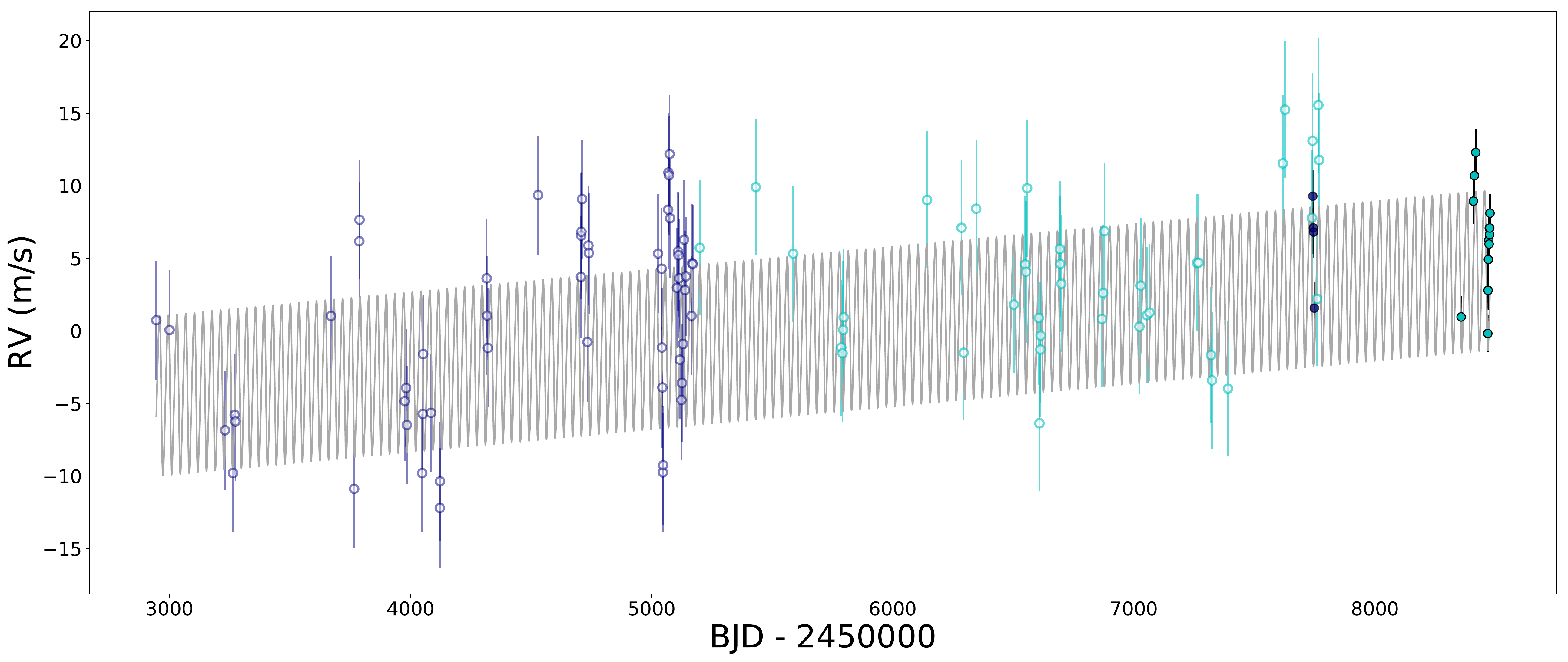}
    \includegraphics[width=2\columnwidth]{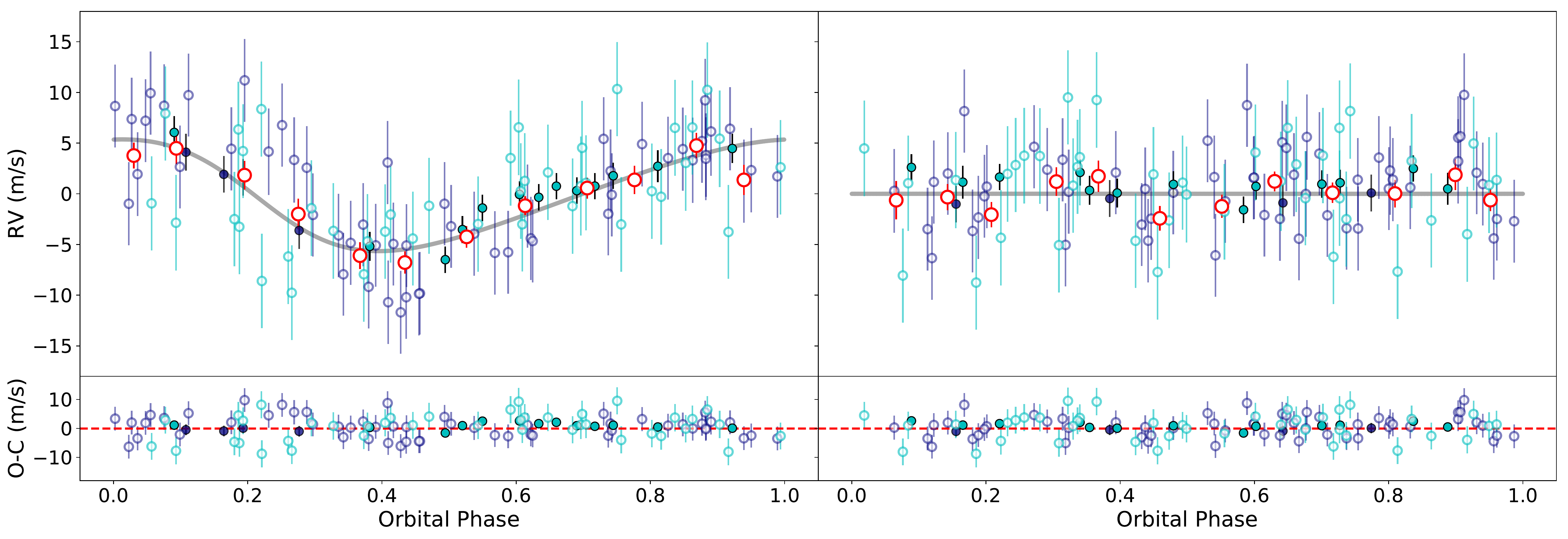}
    \caption{Relative RV measurements and best-fit models for HD 21749. {\it Top:} Complete time series including HARPS pre-upgrade (dark blue open circles), HARPS post-upgrade (dark blue points), PFS pre-upgrade (cyan open circles), and PFS post-upgrade (cyan points) data. For all RVs, the error bars are the quadrature sum of the instrument jitter terms and the measurement uncertainties. The best-fit constant offsets have been subtracted, and the gray line shows the best-fit 2-planet RV model. {\it Bottom:} Phase-folded radial velocities for planet b (left) and planet c (right), with residuals shown below each plot. Point colors are as in the top panel, with the addition of red open circles showing the average velocities binned in 0.08 intervals of orbital phase. Each planet's best-fit model is shown with a gray line, with the Keplerian orbital model for the other planet and the long term trend subtracted.}
    \label{fig:RV}
\end{figure*}

The best-fit parameter values are consistent between our two independent analyses. To make it easier to reproduce our results, we report in this work the values from the \exofast fit (see Table \ref{tab:pars}). The transit and RV observations for each planet, together with the best-fit models, are shown in Figures \ref{fig:tess} and \ref{fig:RV}, respectively. We do not measure a statistically significant RV semi-amplitude for planet c, on which we instead set a 3$\sigma$ upper limit of $\sim$1.43 m/s.

An additional independent analysis of the RV measurements using RadVel \citep{Ful18} with the periods fixed to 35.608 and 7.7882 days and $T_c$ for each planet fixed to the \textit{TESS}-determined values, gives results for $K$, $e$ and $\omega$ that are fully consistent with those in Table \ref{tab:pars}.

Finally, as an experiment before the sectors 3 and 4 data became available, we performed a joint analysis of the sector 1 transit together with the HARPS RVs. The period determination of this fit was driven by the RVs, and the transit only poorly constrained the time of inferior conjunction because of the long baseline between the RV and \tess measurements. Notably, the fit preferred a higher eccentricity (which is no longer allowed when including the second and third transits, and the PFS RVs), and a time of periastron passage offset by 10 days from the true value. The best-fit period was also shorter and the best-fit mass higher than the values obtained when including all transits. 

Our findings underline the need for at least two transits (not necessarily all from \textit{TESS}) to confirm a \tess planet and determine its properties adequately. A single transit is insufficient for most systems, even if RVs are available.

\subsection{Stellar Activity}

\begin{figure*}[h]
    \centering
    \includegraphics[width=2\columnwidth]{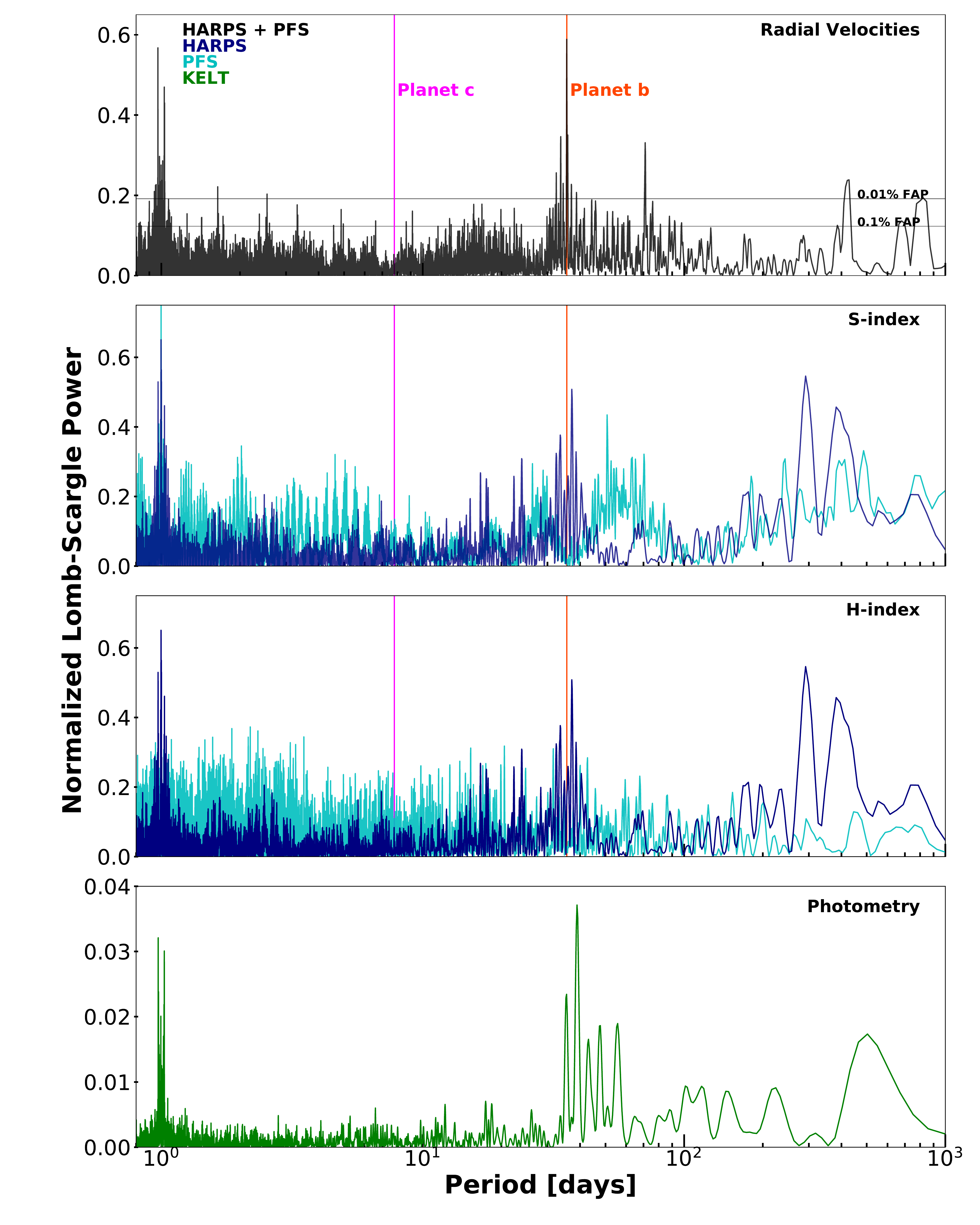}
    \caption{Lomb-Scargle periodograms of, from top to bottom: the complete set of RVs, the $S_{HK}$ index, the H$_{\alpha}$ index and the KELT photometry. Red and magenta lines mark the periods of HD 21749b and HD 21749c, respectively.}
    \label{fig:periodogram}
\end{figure*}

The $log(R_{HK})$ value for HD 21749 suggests a stellar rotation period of 34.5 $\pm$ 7 days \citep{Mam08}. We investigate this further by extracting $S_{HK}$ and H$_{\alpha}$ indices from the PFS (pre-upgrade only) and HARPS spectra. Figure \ref{fig:periodogram} shows Lomb-Scargle periodograms of these indices (second and third panels from the top). In the 10 to 100-day range, the highest peak in both of these HARPS activity indicators corresponds to 37.2 days, which we attribute to the rotation of the star. These peaks also do not overlap with the orbital period of HD 21749b (marked by the red lines).

We also obtained Kilodegree Extemely Little Telescope (KELT; \citealt{Pep2004}) photometry of HD 21749. The star has been monitored by KELT as part of its long-term transit survey of bright stars. The KELT light curve for HD 21749 spans $\sim$3.3 years, contains 7848 individual points (taken between February, 2010 and June, 2013), and has an RMS of $\sim$0.0098 mag. A Lomb-Scargle periodogram of the KELT photometry finds the most significant peak at a period of 38.954 days (see bottom panel of Figure \ref{fig:periodogram}).

More data will help to better determine the stellar rotation period, but the existing photometric and spectroscopic data sets suggest that it is longward of planet b's period. Importantly, the RV periodogram (top panel of Figure \ref{fig:periodogram}) shows the strongest peak at 35.6 days (the period of HD 21749b) but does not show significant (above 0.01\% FAP) power in the 37 to 39-day period range. 

Nevertheless, we employed equation 2 of \cite{Van16b} to estimate the magnitude of systematic errors due to the stellar variability ($\sigma_{s,RV}$) that could affect the RV signals of HD 21749b and HD 21749c. We used the standard deviation of the raw TESS light curve ($\sigma_{s,\tess} = 0.0013$) rather than its peak-to-peak amplitude ($Fpp_{s,\tess}$), and $v$sin$i$ = 1.04 km/s (derived from the HARPS spectra) to obtain $\sigma_{s,RV} \approx 1.3$ m/s.

Therefore, while we do not expect the RV signal of HD 21749b to be strongly affected by stellar variability, the uncertainties on the $K$ values reported in Table \ref{tab:pars} could be somewhat underestimated.

\section{Discussion and conclusion}
\label{sec:conc}

In this Letter we announce the discovery and confirmation of HD 21749b, the second \tess Level-1 planet ($R_P < 4~R_{\oplus}$ and a measured mass) to date, and the longest-period \tess planet confirmed so far. We also report the discovery of HD 21749c, {\it TESS}' first Earth-sized planet.

HD 21749b is a 2.61$^{+0.17}_{-0.16} R_{\oplus}$ planet. Using HARPS and PFS RVs, we measured a mass of 22.7$^{+2.2}_{-1.9} M_\oplus$. Its density of 7.0$^{+1.6}_{-1.3}$ g/cm$^3$ makes it one of the two densest planets with a mass above 15 $M_\oplus$. The other is K2-66b ($M_P$ of $21.3 \pm 3.6 M_\oplus$, $\rho_p$ of $7.8 \pm 2.7$ g/cm$^3$; \citealt{Sin2017}). K2-66b has a density consistent with a rocky composition, but it is also less massive than HD 21749b. On the other hand, the density of HD 21749b indicates it is likely surrounded by a substantial atmosphere. By measuring the density of these two planets (and other similar planets that \tess may find) more precisely, we can begin to observationally constrain the maximum core mass a planet can reach during its formation before accreting a volatile envelope.

According to the transmission spectroscopy metric (TSM) of \citep{Kem18}, HD 21749b is not an ideal target for atmospheric characterization with JWST. However, the TSM is based on 10 hours of JWST observing time. {\it TESS} sub-Neptunes transiting bright K dwarfs and with relatively low equilibrium temperature may be scarce. Therefore, HD 21749b could easily warrant more observing time, particularly to search for species expected in the atmospheres of cooler planets (i.e. methane). It may also be possible to measure HD 21749b's spin/orbit obliquity via the Rossiter McLaughlin effect with either ESPRESSO \citep{Pepe2010} or precise spectrographs planned for the Extremely Large Telescopes.

HD 21749c is an Earth-sized ($R_p = 0.892^{+0.064}_{-0.058} R_{\oplus}$) planet. Its mass is expected to be $\sim$2.5~$M_{\oplus}$ according to the mass-radius relations of \cite{Ning2018} (though the two approximately Earth-sized planets for which we have well-constrained masses so far -- Venus and Earth -- have masses of 1 $M_{\oplus}$ or less). This would give rise to a RV semi-amplitude of $\sim$1.0 m/s, a challenging measurement for the RV spectrographs presented here, but perhaps a feasible measurement with a dedicated VLT/ESPRESSO campaign.

Finally, we emphasize that this work used mainly existing spectroscopic and photometric data, publicly available (as for \citealt{Hua18} and \citealt{Van19}) or generously contributed by multiple ongoing surveys, to confirm and characterize the planets presented in this paper. We postulate that this approach has the potential to be successful for other \tess planet candidates as well.\\

\acknowledgements

We thank the referee for their careful report on the manuscript. The implementation of their suggestions has significantly improved the clarity of the paper. Funding for the \tess mission is provided by NASA's Science Mission directorate. We acknowledge the use of public \tess Alert data from pipelines at the \tess Science Office and at the \tess Science Processing Operations Center. Resources supporting this work were provided by the NASA High-End Computing (HEC) Program through the NASA Advanced Supercomputing (NAS) Division at Ames Research Center for the production of the SPOC data products. This paper includes data collected by the \tess mission, which are publicly available from the Mikulski Archive for Space Telescopes (MAST). This paper is based in part on observations collected at the European Southern Observatory under ESO programmes 076.C-0762(A), 096.C-0499(A), 183.C-0972(A) and 072.C-0488(E). This work makes use of observations from the LCOGT network. DD and JT acknowledge support for this work provided by NASA through Hubble Fellowship grants HST-HF2-51372.001-A and HST-HF2-51399.001-A awarded by the Space Telescope Science Institute, which is operated by the Association of Universities for Research in Astronomy, Inc., for NASA, under contract NAS5-26555. 
MNG, JB, and CXH acknowledge support from MIT’s Kavli Institute as Torres postdoctoral fellows. AV's work was performed under contract with the California Institute of Technology (Caltech)/Jet Propulsion Laboratory (JPL) funded by NASA through the Sagan Fellowship Program executed by the NASA Exoplanet Science Institute. JER was supported by the Harvard Future Faculty Leaders Postdoctoral fellowship. XD acknowledges support from the Branco Weiss Felowship--Society in Science.

\software{EXOFASTv2 \citep{Eas17}, allesfitter (Gunther $\&$ Daylan, in prep.), celerite \citep{Foreman-Mackey2017}, ellc \citep{Max16}, dynesty (\url{https://github.com/joshspeagle/dynesty}), emcee \citep{DFM13}, celerite \citep{Foreman-Mackey2017}, RadVel \citep{Ful18}, SpecMatch \citep{Pet17}}

\facilities{\tess, ESO 3.6 m: HARPS, Magellan Clay: Planet Finder Spectrograph, VLT: NACO, KELT, LCO: NRES, Euler 1.2m: CORALIE}

\bibliographystyle{aasjournal}
\bibliography{research}

\newpage

\begin{deluxetable*}{lccc}
\renewcommand{\arraystretch}{0.9} 
\setlength{\tabcolsep}{0pt} 
\tablewidth{0pt}
\tablecaption{\label{tab:pars} Median values and uncertainties for the HD 21749 system}
\tablehead{\colhead{~~~~~~~~~~~~~~\textbf{Parameter}} &   &  \colhead{~~~~~~~~~~~~~~~~~~~~~~~~~~~~~~~~~~\textbf{Value}}}
\startdata
\smallskip\\\multicolumn{2}{l}{\textbf{Catalogue Stellar Information:}}&\smallskip\\
~~~~R.A.\dotfill & Right Ascension (h:m:s; J2015.5)\dotfill    & ~~~~~~~~~~~~~~~~~~~~~~~~~~~~~~~~~~03:27:00.045 \\
~~~~Dec\dotfill & Declination (d:m:s; J2015.5)\dotfill & ~~~~~~~~~~~~~~~~~~~~~~~~~~~~~~~~~~-63:30:00.60\\
~~~~$\lambda$ \dotfill & Ecliptic longitude (deg) \dotfill & ~~~~~~~~~~~~~~~~~~~~~~~~~~~~~~~~~~352.8828 \\
~~~~$\beta$ \dotfill & Ecliptic latitude (deg) \dotfill &~~~~~~~~~~~~~~~~~~~~~~~~~~~~~~~~~~-73.8346\\
~~~~HD ID\dotfill & Henry Draper Catalogue ID \dotfill & ~~~~~~~~~~~~~~~~~~~~~~~~~~~~~~~~~~21749\\
~~~~TIC ID\dotfill & \tess Input Catalogue ID \dotfill & ~~~~~~~~~~~~~~~~~~~~~~~~~~~~~~~~~~279741379\\
~~~~TOI ID\dotfill & \tess Object of Interest ID \dotfill & ~~~~~~~~~~~~~~~~~~~~~~~~~~~~~~~~~~186.01
\smallskip\\\multicolumn{2}{l}{\textbf{Photometric Stellar Properties:}}&\smallskip\\
~~~~V mag\dotfill & Apparent V-band magnitude \dotfill & ~~~~~~~~~~~~~~~~~~~~~~~~~~~~~~~~~~8.1\\
~~~~\tess mag\dotfill & Apparent \tess-band magnitude \dotfill &~~~~~~~~~~~~~~~~~~~~~~~~~~~~~~~~~~6.95
\smallskip\\\multicolumn{2}{l}{\textbf{Stellar Parameters:}}&\smallskip\\
~~~~$D$ \dotfill &Distance (pc)\dotfill & ~~~~~~~~~~~~~~~~~~~~~~~~~~~~~~~~~~$16.33 \pm 0.007$ \\
~~~~$M_*$\dotfill &Mass (\msun)\dotfill &~~~~~~~~~~~~~~~~~~~~~~~~~~~~~~~~~~$0.73 \pm 0.07$\\
~~~~$R_*$\dotfill &Radius (\rsun)\dotfill &~~~~~~~~~~~~~~~~~~~~~~~~~~~~~~~~~~$0.695 \pm 0.030$\\
~~~~$L_*$\dotfill &Luminosity (\lsun)\dotfill &~~~~~~~~~~~~~~~~~~~~~~~~~~~~~~~~~~$0.20597 \pm 0.00016$\\
~~~~$\rho_*$\dotfill &Density (cgs)\dotfill &~~~~~~~~~~~~~~~~~~~~~~~~~~~~~~~~~~$3.03^{+0.50}_{-0.47}$\\
~~~~$\log{g}$\dotfill &Surface gravity (cgs)\dotfill &~~~~~~~~~~~~~~~~~~~~~~~~~~~~~~~~~~$4.613^{+0.052}_{-0.061}$\\
~~~~$T_{\rm eff}$\dotfill &Effective Temperature (K)\dotfill &~~~~~~~~~~~~~~~~~~~~~~~~~~~~~~~~~~$4640 \pm 100$\\
~~~~$[{\rm Fe/H}]$\dotfill &Metallicity (dex)\dotfill &~~~~~~~~~~~~~~~~~~~~~~~~~~~~~~~~~~$0.003\pm0.060$ \\
~~~~$Age$ \dotfill &Stellar age (Gyr) \dotfill &~~~~~~~~~~~~~~~~~~~~~~~~~~~~~~~~~~$3.8 \pm 3.7$
\smallskip\\\multicolumn{2}{l}{\textbf{MCMC Fit \tess Bandpass Wavelength Parameters:}}&\smallskip\\
~~~~$u_{1}$\dotfill &linear limb-darkening coeff \dotfill &~~~~~~~~~~~~~~~~~~~~~~~~~~~~~~~~~~$0.492\pm0.026$\\
~~~~$u_{2}$\dotfill &quadratic limb-darkening coeff \dotfill &~~~~~~~~~~~~~~~~~~~~~~~~~~~~~~~~~~$0.170^{+0.026}_{-0.025}$\\
~~~~$A_D^a$\dotfill &Dilution from neighboring stars \dotfill &~~~~~~~~~~~~~~~~~~~~~~~~~~~~~~~~~~$0.121\pm0.012$
\smallskip\\\multicolumn{2}{l}{\textbf{MCMC Fit Telescope Parameters:}}&\smallskip\\
~~~~$\dot{\gamma}$\dotfill &RV slope (m/s/day)\dotfill &~~~~~~~~~~~~~~~~~~~~~~~~~~~~~~~~~~$0.00157^{+0.00070}_{-0.00068}$\\
~~~~$\gamma_{\rm rel,1}$\dotfill &Relative RV Offset HARPS 1 (m/s)\dotfill &~~~~~~~~~~~~~~~~~~~~~~~~~~~~~~~~~~$59608.99^{+0.66}_{-0.64}$\\
~~~~$\gamma_{\rm rel,2}$\dotfill &Relative RV Offset HARPS 2 (m/s)\dotfill &~~~~~~~~~~~~~~~~~~~~~~~~~~~~~~~~~~$59619.1^{+3.2}_{-3.3}$\\
~~~~$\gamma_{\rm rel,3}$\dotfill &Relative RV Offset PFS 1 (m/s)\dotfill &~~~~~~~~~~~~~~~~~~~~~~~~~~~~~~~~~~$-4.6\pm2.0$\\
~~~~$\gamma_{\rm rel,4}$\dotfill &Relative RV Offset PFS 2 (m/s)\dotfill &~~~~~~~~~~~~~~~~~~~~~~~~~~~~~~~~~~$-5.3^{+3.0}_{-3.1}$\\
~~~~$\sigma_{J,1}$\dotfill &RV Jitter HARPS 1 (m/s)\dotfill &~~~~~~~~~~~~~~~~~~~~~~~~~~~~~~~~~~$4.17^{+0.47}_{-0.41}$\\
~~~~$\sigma_{J,2}$\dotfill &RV Jitter HARPS 2 (m/s)\dotfill &~~~~~~~~~~~~~~~~~~~~~~~~~~~~~~~~~~$1.7^{+4.9}_{-1.7}$\\
~~~~$\sigma_{J,3}$\dotfill &RV Jitter PFS 1 (m/s)\dotfill &~~~~~~~~~~~~~~~~~~~~~~~~~~~~~~~~~~$4.63^{+0.65}_{-0.53}$ \\
~~~~$\sigma_{J,4}$\dotfill &RV Jitter PFS 2 (m/s)\dotfill &~~~~~~~~~~~~~~~~~~~~~~~~~~~~~~~~~~$1.27^{+0.42}_{-0.29}$
\smallskip\\ & & \textbf{HD 21749b} &  \textbf{HD 21749c}\\
\multicolumn{2}{l}{\textbf{MCMC Fit Planetary Parameters:}}&\smallskip\\
~~~~$P$\dotfill &Period (days)\dotfill &$35.61253^{+0.00060}_{-0.00062}$&$7.78993^{+0.00051}_{-0.00044}$\\
~~~~$T_0$\dotfill &Optimal conjunction Time (\bjdtdb)\dotfill &$2458385.92502^{+0.00054}_{-0.00055}$&$2458371.2287^{+0.0016}_{-0.0015}$\\
~~~~$i$\dotfill &Inclination (Degrees)\dotfill &$89.33^{+0.15}_{-0.11}$&$88.90^{+0.50}_{-0.37}$\\
~~~~$R_P/R_*$\dotfill &Radius of planet in stellar radii \dotfill &$0.0350^{+0.0011}_{-0.0010}$&$0.01196^{+0.00051}_{-0.00052}$\\
~~~~$K$\dotfill &RV semi-amplitude (m/s)\dotfill &$5.51^{+0.41}_{-0.33}$&$< 1.43^b$\\
~~~~$ecos{\omega_*}$\dotfill & \dotfill &$-0.025\pm0.051$&--\\
~~~~$esin{\omega_*}$\dotfill & \dotfill &$0.179^{+0.078}_{-0.085}$&--\\
\smallskip\\\multicolumn{2}{l}{\textbf{Derived Planetary Parameters:}}&\smallskip\\
~~~~$a/R_*$\dotfill &Semi-major axis in stellar radii \dotfill &$60.1^{+3.2}_{-3.3}$&$21.8\pm1.2$\\
~~~~$R_P$\dotfill &Radius ($R_{\oplus}$)\dotfill &$2.61^{+0.17}_{-0.16}$&$0.892^{+0.064}_{-0.058}$\\
~~~~$a$\dotfill &Semi-major axis (AU)\dotfill &$0.1915^{+0.0058}_{-0.0063}$&$0.0695^{+0.0021}_{-0.0023}$\\
~~~~$T_{eq}$\dotfill &Equilibrium temperature (K)\dotfill &$422^{+15}_{-14}$&$701^{+25}_{-23}$\\
~~~~$e$\dotfill &Eccentricity \dotfill &$0.188^{+0.076}_{-0.078}$&--\\
~~~~$\omega_*$\dotfill &Argument of Periastron (Degrees)\dotfill &$98^{+21}_{-17}$&--\\
~~~~$T_P$\dotfill &Time of Periastron (\bjdtdb)\dotfill &$2458350.8^{+1.6}_{-1.1}$&$2458332.2789^{+0.0026}_{-0.0027}$\\
~~~~$M_P$\dotfill &Mass ($M_{\oplus}$)\dotfill &$22.7^{+2.2}_{-1.9}$&$< 3.70^b$\\
~~~~$\tau$\dotfill &Ingress/egress transit duration (hours)\dotfill &$0.163^{+0.043}_{-0.034}$&$0.0358^{+0.0055}_{-0.0046}$\\
~~~~$T_{14}$\dotfill &Total transit duration (hours)\dotfill & $3.230^{+0.041}_{-0.038}$&$2.515^{+0.077}_{-0.086}$\\
~~~~$b$\dotfill &Transit Impact parameter \dotfill &$0.587^{+0.097}_{-0.15}$&$0.42^{+0.11}_{-0.18}$\\
~~~~$\rho_P$\dotfill &Density (cgs)\dotfill &$7.0^{+1.6}_{-1.3}$&$< 31.93^b$\\
~~~~$logg_P$\dotfill &Surface gravity \dotfill &$3.514^{+0.067}_{-0.069}$&$< 3.68^b$\\
\enddata
\tablenotetext{a}{$A_D$ is defined as $F2/(F1+F2)$, where $F2$ is the flux of the target and $F1$ is the flux of the neighboring contaminating star.}
\tablenotetext{b}{Limits denote the 99.73\% (3$\sigma$) upper confidence interval.}
\end{deluxetable*}

\end{document}